\documentclass{ametsocV6.1}

\title{Bridging the Sensitivity Gap in Precipitation Estimates from Spaceborne Radars using Passive Microwave Observations}

\usepackage{booktabs}

\authors{Simon Pfreundschuh,\aff{a}\correspondingauthor{Simon Pfreundschuh, simon.pfreundschuh@colostate.edu}
Christian D. Kummerow,\aff{a}
}

\affiliation{\aff{a}{Department of Atmospheric Science, Colorado State University, Fort Collins, USA}}

\abstract{%
Current global precipitation estimates from spaceborne precipitation radars are
limited by their sensitivity to light and frozen precipitation, leading to
systematic underestimation of precipitation at high latitudes. Because
passive microwave retrievals are commonly trained using these radar observations
as reference data, this limitation is propagated into passive microwave
precipitation products.\\
This study introduces a novel passive microwave oceanic precipitation retrieval,
GPROF-NN eXtended Precipitation Regime (XPR), that combines reference estimates
from a cloud radar and a precipitation radar to overcome the sensitivity
limitations of current spaceborne precipitation radars. The retrieval is trained
to estimate light precipitation from CloudSat observations and moderate-to-heavy
precipitation using observations from the GPM Dual-Frequency Precipitation
Radar. The two estimates are combined using a fusion scheme to obtain a
consistent precipitation estimate across precipitation regimes. Validation
against in situ measurements from shipborne disdrometers shows a 26\%
improvement in the detection skill for high-latitude precipitation in terms of
the critical success index and a reduction in the underestimation of
high-latitude and frozen precipitation by more than 50\% compared to retrievals
constrained only by precipitation radar data. However, the fused retrieval does
not improve the precision of instantaneous precipitation estimates, which is
likely due to significant random errors in the CloudSat-based reference
estimates of liquid precipitation.\\
These results demonstrate that passive-microwave retrievals can leverage the
complementary sensitivities of cloud and precipitation radars to provide more
consistent precipitation estimates across precipitation regimes than either
reference instrument alone. The proposed retrieval provides a pathway to improve
the representation of oceanic precipitation in future GPM precipitation
products.\\
}

\begin{document}

%% Necessary!
\maketitle

%%%%%%%%%%%%%%%%%%%%%%%%%%%%%%%%%%%%%%%%%%%%%%%%%%%%%%%%%%%%%%%%%%%%%
% SIGNIFICANCE STATEMENT/CAPSULE SUMMARY
%%%%%%%%%%%%%%%%%%%%%%%%%%%%%%%%%%%%%%%%%%%%%%%%%%%%%%%%%%%%%%%%%%%%%
%
% If you are including an optional significance statement for a journal article or a required capsule summary for BAMS 
% (see www.ametsoc.org/ams/index.cfm/publications/authors/journal-and-bams-authors/formatting-and-manuscript-components for details), 
% please apply the necessary command as shown below:
%
% Significance Statement (all journals except BAMS)
%
\statement

    Current satellite-based precipitation datasets underestimate high-latitude
    oceanic precipitation because the radar observations they are based on are
    insensitive to light and frozen precipitation. We address this issue by
    developing a novel precipitation estimation algorithm that uses passive
    microwave observations to estimate precipitation based on reference data
    from two different types of spaceborne radars: a cloud radar sensitive to
    light precipitation and a precipitation radar sensitive to moderate to heavy
    precipitation. This novel approach significantly improves precipitation
    detection capability and reduces systematic underestimation of oceanic
    precipitation at high latitudes, which we show by extensive validation
    against independent precipitation measurements from shipborne instruments.
    While the results demonstrate that global precipitation estimates of oceanic
    precipitation can be improved by leveraging the complementary observation
    capabilities of cloud and precipitation radars, additional analysis of the
    radar based estimates also reveals inconsistencies between precipitation
    estimates from the two radars highlighting remaining shortcomings
    precipitation estimates from spaceborne radars.

    The algorithm developed in this work will be used to fill in missing light and
    frozen precipitation in the next version of passive microwave precipitation
    datasets from NASA's Global Precipitation Measurements mission. It presents
    an important step towards improved observations of the hydrological cycle at
    high latitudes.

%%%%%%%%%%%%%%%%%%%%%%%%%%%%%%%%%%%%%%%%%%%%%%%%%%%%%%%%%%%%%%%%%%%%%
% MAIN BODY OF PAPER
%%%%%%%%%%%%%%%%%%%%%%%%%%%%%%%%%%%%%%%%%%%%%%%%%%%%%%%%%%%%%%%%%%%%%
%
\section{Introduction}

Precipitation is an important atmospheric process critical for sustaining
terrestrial ecosystems and affecting a wide range of societal and economic
activities. Measuring the global distribution of precipitation and its
variability is an important prerequisite towards understanding – and,
ultimately, predicting – the occurrence and intensity of precipitation.
Satellite remote sensing is the principal technique for monitoring the global
distribution of precipitation and therefore an essential tool for advancing the
understanding of the global hydrological cycle.

\begin{figure}[h]
 \centerline{\includegraphics[width=37pc]{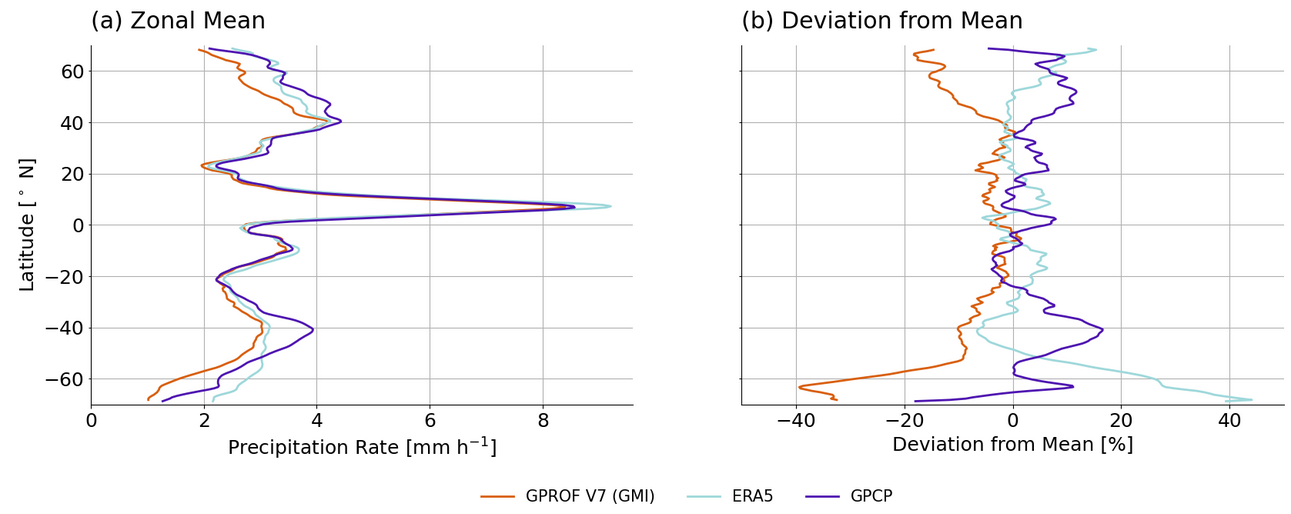}}
  \caption{Zonal means of oceanic precipitation from GMI, the GPCP merged precipitation dataset, and the ERA5 reanalysis. Panel (a) shows absolute precipitation rates; Panel (b) shows relative deviations from the cross-dataset mean.}\label{fig:zonal_means}
\end{figure}

However, even after three decades of dedicated satellite missions and scientific
efforts to measure global precipitation, oceanic precipitation at high latitudes still
remains poorly constrained. To illustrate this, Fig. \ref{fig:zonal_means} shows zonal means of
oceanic precipitation from three principal precipitation datasets: passive microwave (PMW)
precipitation estimates from GPM Microwave Imager (GMI) retrieved using  V07
of the Goddard Profiling Algorithm (GPROF, \citet{kummerow_evolution_2015}) retrieval algorithm, the Global
Precipitation Climatology Project (GPCP, \citet{huffman_new_2023}), and the ERA5
reanalysis \citep{hersbach_era5_2020}. While the datasets are largely consistent
across the tropics and mid-latitudes, the zonal means diverge poleward of 40 $^\circ$
showing deviations of up to 80 \% in the mid and high latitudes.

These large differences are due to a number of challenges. Light and frozen
precipitation (< 0.1 mm/hr) contribute relatively more to total precipitation
accumulations at high latitudes than in the tropics \citep{milani_state_2023},
and are harder to simulate in models due to weaker linkages to synoptic
forcings. Light and frozen precipitation are also harder to estimate from
satellite observations due to their weaker radiometric signatures
\citep{kidd_global_2011} and larger dependence on microphysical variations
\citep{ekelund_2020}. In addition to this, current PMW retrieval algorithms such
as GPROF are constructed to reproduce reference precipitation from the GPM
Dual-Frequency Precipitation Radar (DPR). The limited sensitivity of the DPR
causes these reference precipitation estimates to miss shallow and frozen
precipitation and this behavior is in turn reproduced by the PMW retrieval
algorithms based on the DPR reference estimates.

The CloudSat Cloud Profiling Radar (CPR, \citet{stephens2008cloudsat}) is a different
spaceborne radar designed primarily to observe clouds. It operates at a frequency of 94.05
GHz, which makes it more sensitive to light precipitation and snowfall. The
higher sensitivity,its higher spatial resolution and lower surface
clutter height significantly improve its capability to detect precipitation compared to
the DPR. However, the observations are also heavily attenuated by liquid precipitation,
which complicates deriving quantitative precipitation estimates particularly for moderate
and heavy precipitation.

Merged precipitation products such as the Integrated Multi-Satellite Retrieval
for the Global Precipitation Measurement Mission (IMERG,
\citet{huffman_integrated_2020}) and GPCP that rely on PMW retrievals compensate
for the inherited insensitivity to light and frozen precipitation using
climatological adjustments. GPCP, for example, rescales microwave retrievals
using precipitation climatologies derived in part from the CPR, wich allows
extending reference observations to higher latitudes in addition to providing
higher sensitivity to light and frozen precipitation. However, such corrections
constrain long-term means rather than instantaneous retrievals and may therefore
misrepresent the temporal variability of high-latitude precipitation.

Recent neural-network precipitation retrievals \citep{pfreundschuh_gprof_2024}
have improved performance across many regimes and are being adopted
operationally. However, machine-learning retrievals approximate the conditional
expectation defined by their training data and thus reproduce biases present in
the reference observations. Training solely on DPR-based precipitation therefore
propagates the radar’s limited sensitivity to light and frozen precipitation
into the learned retrieval, regardless of retrieval technique used.

Here we aim to overcome the limitations of current DPR-based precipitation
estimates by using PMW observations to bridge the sensitivity gap between the
CPR and DPR precipitation estimates and thus achieve more complete
representation of precipitation across regimes. Specifically, we propose
GPROF-NN eXtended Precipitation Regime (XPR), a GMI-based retrieval that jointly
predicts precipitation estimates based on CPR- and DPR-based reference
precipitation rates. Since the CPR operates at a higher frequency, it is more
sensitive to small and frozen hydrometeors and can thus be expected to provide a
better representation of light and frozen precipitation. DPR, on the other hand,
misses light and frozen precipitation but provides reliable estimates of
moderate and heavy precipitation. The GPROF-NN XPR retrieval is trained to map
GMI PMW observations to both a light precipitation estimate based on CPR
reference data and an estimate of moderate to heavy precipitation based on DPR
reference data. As we will show here, an improved representation of
precipitation across regimes can be obtained by using the CPR-based
precipitation estimates to fill in missing light precipitation in the DPR-based
Estimates.

GPROF-NN XPR has been developed as an auxiliary retrieval to recover missing
oceanic precipitation in DPR-based reference data used by previous versions of
the GPROF and GPROF-NN \citep{pfreundschuh_gprof-nn_2022} retrievals. It is part
of the development of GPROF V08 -- the next operational version of the GPM PMW
retrievals -- and aims to address the systematic underestimation of high
latitude precipitation. The previous operational version of GPROF -- GPROF V07
-- used light precipitation estimates from the MiRS algorithm
\citep{boukabara_mirs_2011} to fill in missing precipitation in DPR estimates
but this was unsuccessful in fully addressing the underestimation. The filling
in of missing oceanic precipitation complements several other interventions
addressed to fix shortcomings in the DPR data used as a priori database and
training dataset such as the use of ground-based radar estimates over
snow-covered surface and the use of ERA5 precipitation over sea ice.

This manuscript describes the GPROF-NN XPR retrieval and validates its
precipitation estimates using in-situ measurements from shipborne distrometers
demonstrating its ability to improve the detection of light and frozen
precipitation and reduce  climatological biases. Additionally, we
provide a validation of the CPR- and DPR-based reference estimates against
independent measurements from ground-based precipitation radars.
Section~\ref{sec:methods} describes the retrieval implementation as well as the
data used to validate the retrievals. Section~\ref{sec:results} assesses and
validates the GPROF-NN XPR retrieval, validates the radar-based reference
estimates, and assesses the effect of the enhanced sensitivity of the GPROF-NN
XPR retrieval on the climatological precipitation distribution. Finally,
section~\ref{sec:summary} summarizes and discusses the main findings from this work.

\section{Methods and Data}
\label{sec:methods}

The GPROF-NN XPR retrieval algorithm is a machine-learning based precipitation
retrieval that is trained to simultaneously provide estimates of CPR-based and
DPR-based precipitation rates. The two estimates are subsequently fused to
provide a general precipitation estimate that better represents precipitation
across the full spectrum of intensities. Below we describe the retrieval
implementation together with the data sources used to validate the retrieval
itself as well as the two radar-based reference datasets.

\subsection{The GPROF-NN XPR Retrieval}

The GPROF-NN XPR retrieval is based on the GPROF-NN retrieval framework
introduced in \citep{pfreundschuh_gprof-nn_2022} and subsequently validated in
\citep{pfreundschuh_gprof_2024}. It uses the same neural-network architecture as
the upcoming version of GPROF-NN, which is currently under development and will
become the operational precipitation retrieval of the GPM constellation as GPROF
V08.

\subsubsection*{Neural Network Model}

The GPROF-NN XPR retrieval employs a U-Net–type encoder–decoder architecture
composed of EfficientNet-V2–style inverted bottleneck convolutional blocks
\citep{tan2021efficientnetv2}. Following the normalization and activation design
adopted in ConvNeXt \citep{liu2022convnet}, each convolutional block uses Layer
Normalization \citep{ba2016layer} and GELU activation functions
\citep{hendrycks2016gaussian}. The network input consists of GMI brightness
temperatures together with the corresponding earth-incidence angles (EIA) for
all channels and a set of ancillary variables. These ancillary data describe
both surface conditions (land fraction, elevation, mountain classification,
leaf-area index, ice fraction, and snow depth) and the meteorological
environment (2-m temperature, 10-m wind speed, orographic wind, and total column
water vapor).

The model produces two precipitation estimates: one trained on CPR-based
reference precipitation and one trained on DPR-based reference precipitation.
Both estimates are derived from a shared latent representation using separate
multilayer perceptron output heads. Each head predicts 32 quantiles of the
conditional precipitation distribution, which are learned using quantile
regression \citep{pfreundschuh_neural_2018}. The expectation is that the
CPR-based reference data will miss heavy rain where CPR is attenuated while the
DPR-based reference will miss light precipitation below the detection threshold
of those radars.

\begin{figure}[h]
 \centerline{\includegraphics[width=37pc]{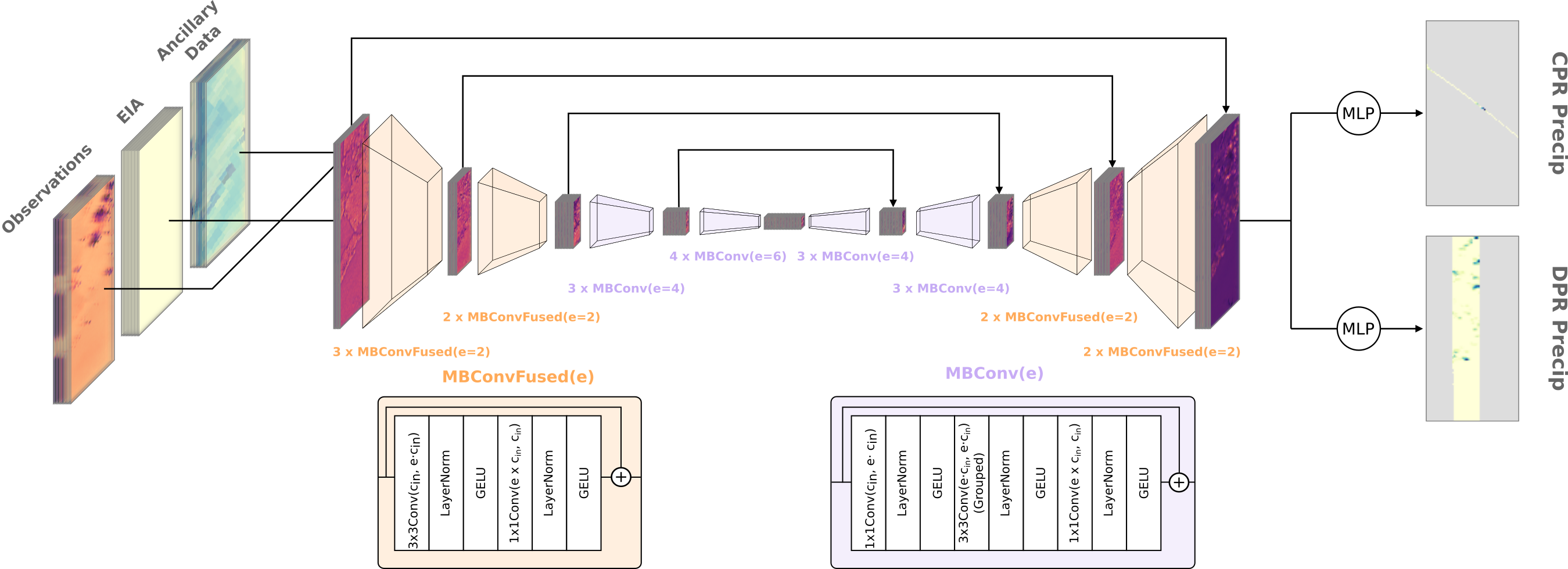}}
 \caption{
   Illustration of the U-Net neural network used by GPROF-NN XPR. The network
   employs MobileNet convolutional blocks (MBConv) within an encoder–decoder
   architecture. Input features consist of image patches of GMI brightness
   temperatures, corresponding earth-incidence angles (EIA), and ancillary
   variables. These inputs are mapped to a shared latent representation from which
   two precipitation estimates are derived through separate output heads: a
   CPR-based light precipitation estimate and a DPR-constrained
   moderate-to-heavy precipitation estimate.
 }\label{fig:neural_network}
\end{figure}

\subsubsection*{Training Data and Training}

The training dataset for the GPROF-NN XPR retrieval is constructed from two
independent sets of collocations between GMI observations and reference
precipitation estimates from the CPR and DPR sensors.

For the GMI--CPR dataset, precipitation estimates are derived by combining
information from the 2C-Precip-Column \citep{haynes_2009}, 2C-Rain-Profile
\citep{lebsock_2011}, and 2C-Snow-Profile \citep{wood_2014} products.
Precipitation rates are assigned according to the precipitation classification
in 2C-Precip-Column: pixels flagged as non-precipitating are assigned zero
precipitation; pixels classified as certain rain are assigned the rate from
2C-Rain-Profile; pixels classified as certain snow are assigned the rate from
2C-SnowProfile. Pixels that do not meet any of these conditions or are assigned
a retrieval confidence less than three are treated as missing and excluded from
training. The dataset includes all GMI-CPR collocations from 2014–2017 with a maximum
temporal separation of 15 min between the GMI and CPR observations.
Collocations from January–October 2018 are used for validation, and those from
October 2018–October 2019 are reserved for testing.

For the GMI–DPR dataset, precipitation rates are taken from the GPM 2BCMB \citep{grecu2016gpm}
combined radar–radiometer product. Because GMI and DPR observe from the same
platform, these collocations are substantially more numerous than the
GMI–CPR matchups. During training, the GMI–CPR samples are therefore
oversampled such that, on average, every second training sample originates from
the GMI–CPR dataset, ensuring similar representation of both reference
regimes. The training is performed across 120 epochs using a cosine-annealing
learning-rate schedule with an initial learning rate of 0.0005 and warm restarts
after 20 and 60 epochs.

\subsubsection*{Fused Precipitation Estimates}

\begin{figure}[h]
 \centerline{\includegraphics[width=37pc]{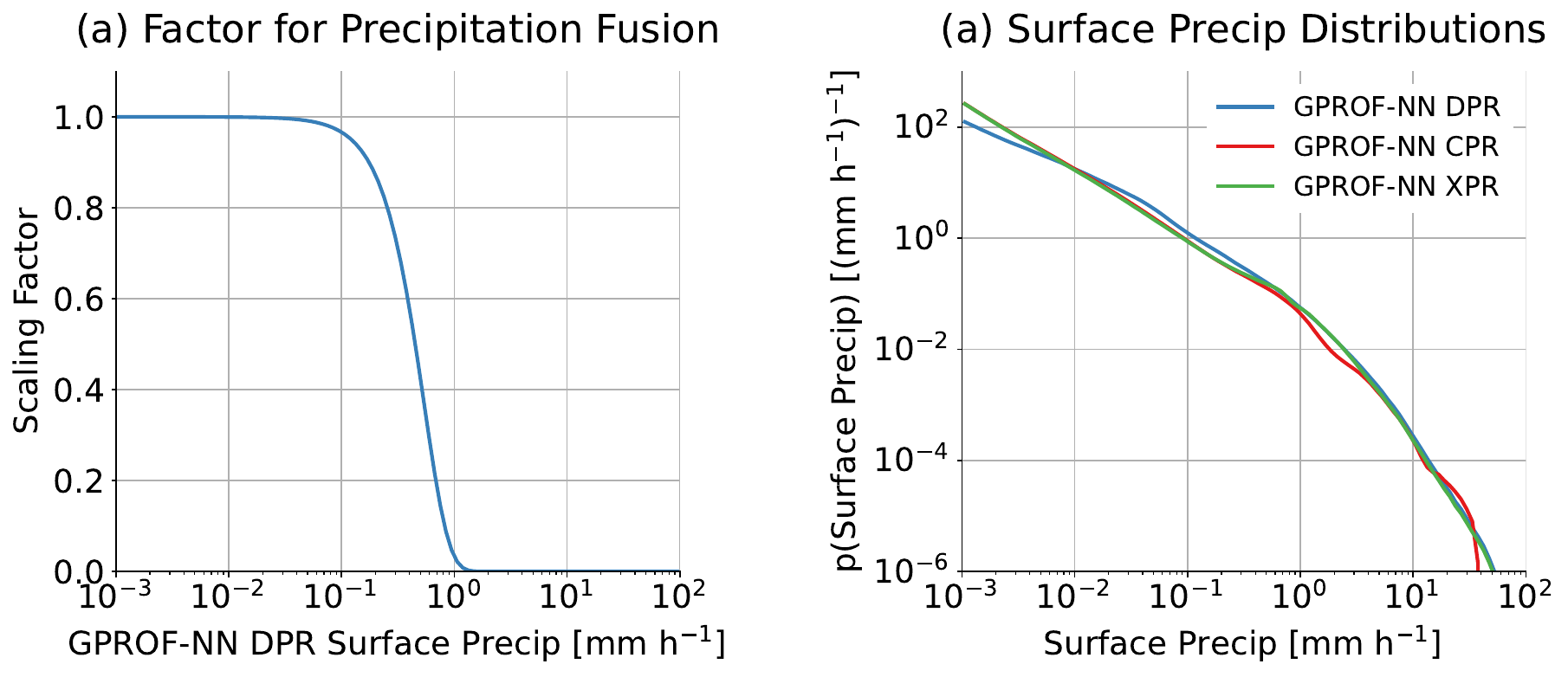}}
 \caption{
Scaling factors used to fuse CPR- and DPR-based precipitation estimates (a) and
resulting precipitation rate distributions (b).
 }\label{fig:fused_distributions}
\end{figure}

The GPROF-NN XPR retrieval produces a CPR-based and a DPR-based precipitation
estimate for each GMI pixel. To provide a single precipitation estimate that
properly describes precipitation across precipitation regimes, the two estimates
must be combined. The fused precipitation relies on the CPR-based estimate when the
DPR-based retrieval indicates no precipitation (reflectivities below the DPR
threshold) and transitions to the DPR-based estimate as intensity increases
(reflectivities likely attenuated in CPR). A simple linear transition, however,
was found to introduce a notable artifact in the distribution of the fused
precipitation rates. Instead, a zero-centered Gaussian weighting function was
applied to obtain a smooth precipitation-rate distribution. The Gaussian has a
full width at half maximum of 0.45 mm h$^{-1}$, with the resulting weighting
function as well as the distribution of test-data precipitation rates shown in
Fig.~\ref{fig:fused_distributions}.

The distribution of the fused precipitation estimates follows that of the
CPR-based estimates at low precipitation rates and transitions to follow the
distribution of the DPR-based precipitation rates for moderate to high
precipitation. The distribution of the GPROF-NN CPR precipitation estimates
exhibits several irregularities and is less smooth than that of the GPROF-NN DPR
estimates. We suspect that this may be related to the bimodal distribution of
the 2C-Rain-Profile estimates that was noted already in Lebsock (2011). While
the bulges in the PDF of GPROF-NN CPR precipitation estimates do not match the
exact peaks of the distribution in Fig. 7 in Lebsock (2011), they may be shifted
in the retrieved distributions due to the increased resolution of the GMI-based
GPROF-NN retrievals.

\subsection{Validation Data}

We validate both the GPROF-NN XPR retrieval and the underlying CPR- and
DPR-based precipitation estimates using independent observations. The GMI-based
GPROF-NN XPR estimates are validated against in situ measurements from shipborne
disdrometers, which provide the most direct reference data available for oceanic
precipitation. Owing to the much narrower swaths of the DPR and CPR, there are
not enough overpasses to meaningfully evaluate the radar-based reference
estimates with the shipborne measurements. Instead, their consistency is
assessed using overpasses along the U.S. East Coast, where the satellite
estimates are compared with coincident observations from ground-based weather
radars.

\subsubsection*{OceanRAIN Ship-Based Distrometer Measurements}

\begin{figure}[h]
 \centerline{\includegraphics[width=37pc]{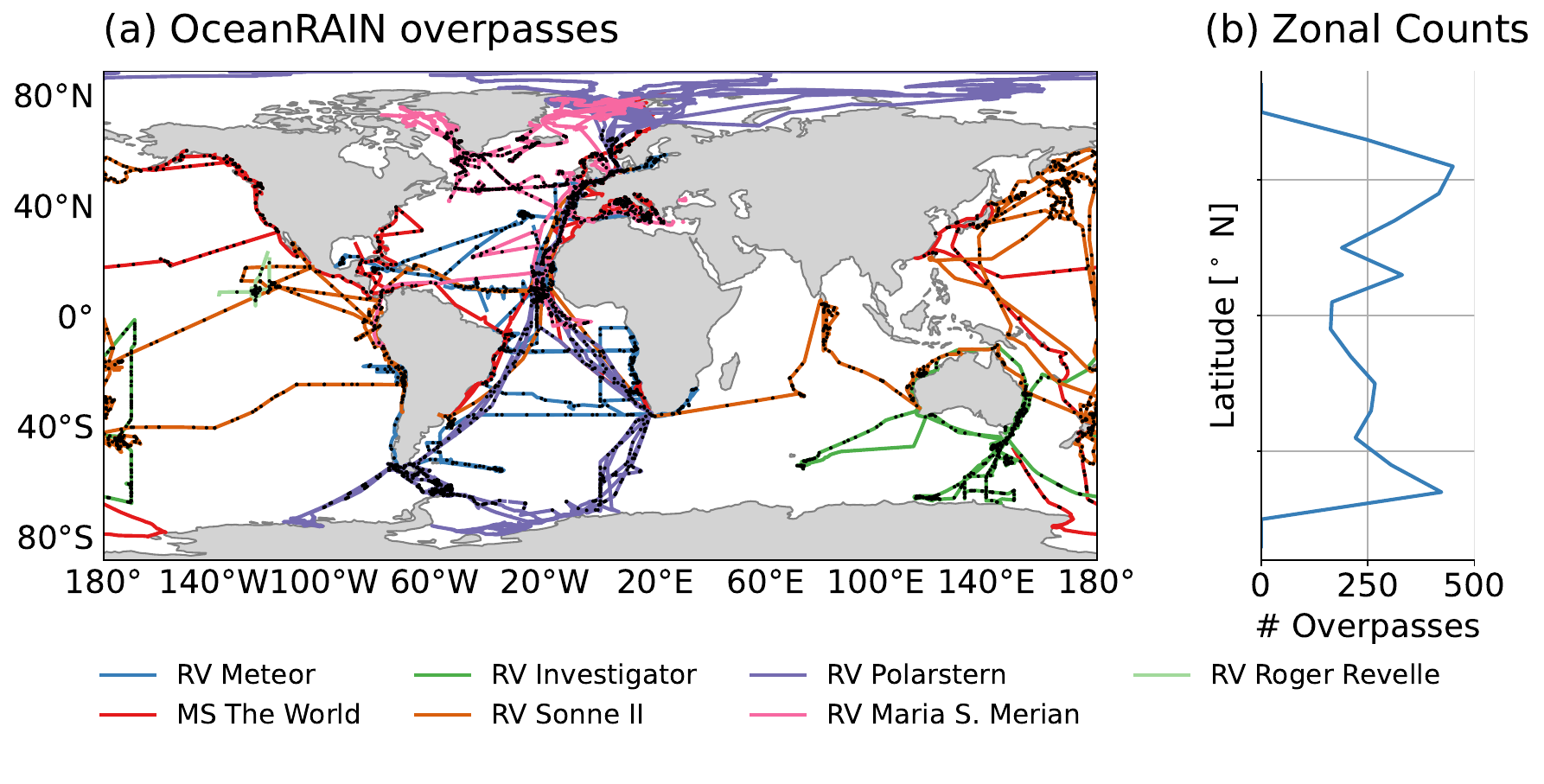}}
 \caption{
OceanRAIN ship tracks and corresponding GMI overpasses. Colored lines indicate
the trajectories of OceanRAIN vessels during the GMI observation period, and
black dots denote the collocated satellite overpasses used for validation of the
GPROF-NN XPR retrieval.
 }\label{fig:ocean_rain_collocs}
\end{figure}

The OceanRAIN dataset \citep{klepp2018oceanrain} provides shipborne disdrometer
measurements of precipitation rates over the ocean. For validation of the
GPROF-NN XPR retrieval, we identify GMI overpasses of the ship locations from
March 2014, the start of GMI observations, through September 2018, prior to the
period used to extract GPM--DPR training collocations. Because this interval overlaps
with the GMI–CPR training dataset, 64 matchups coinciding with CloudSat overpasses
were excluded from the validation dataset.

Figure~\ref{fig:ocean_rain_collocs} shows the ship tracks associated with the OceanRAIN campaigns together
with the corresponding GMI overpasses. To account for the spatial scale mismatch
between satellite footprints and the distrometer measurements, the satellite estimates are
compared with 30-min rolling averages of the disdrometer precipitation rates.
The matched satellite precipitation is computed by computing the  GMI pixel closest to the ship
position for each minute in the averaging window and averaging the resulting precipitation estimates.

To eliminate the impact of errors in the validation data, a relative error is calculated using

\begin{align}
  \text{err} = \frac{|p_\text{retrieved} - p_\text{reference}|}{max(0.5, 0.5 \cdot |p_\text{retrieved} + p_\text{reference}|)}.
\end{align}

The error is calculated for the GPROF-NN DPR retrieval as well as the
corresponding GPROF V07 and ERA5 precipitation estimates. Samples where the
minimum relative error across these three datasets exceeds 1.8 are discarded. In
total, 26 samples from the 3436 collocations with valid precipitation estimates
fail to meet this quality criterion. The removed overpasses correspond to
precipitation events missed by the distrometer as well as cases where the
distrometer indicates very high precipitation rates not seen in the satellite or
ERA5 data. The exclusion of these events does not impact the qualitative results
of the validation.

\subsubsection*{Ground-Based Precipitation Radar Estimates}

\begin{figure}[h]
 \centerline{\includegraphics[width=37pc]{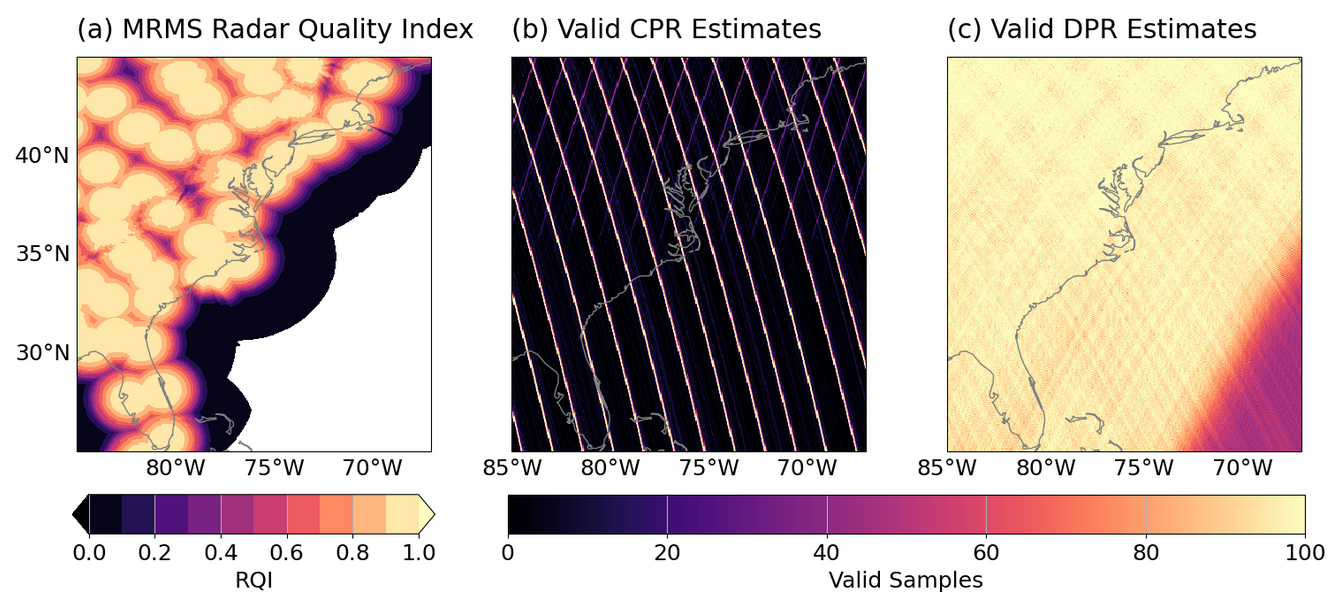}}
 \caption{
Summary statistics the validation data derived from NOAA’s Multi-Radar
Multi-Sensor data. Panel (a) shows the radar-quality index quantifying the
quality of the radar measurements. Panel (b) and (c ) show the mean
precipitation and precipitation occurrence, respectively.
 }\label{fig:radar_overpasses}
\end{figure}

To provide additional context for the evaluation of the GPROF-NN XPR retrieval,
we assess the CPR and DPR reference precipitation estimates using independent
observations from NOAA’s Multi-Radar Multi-Sensor (MRMS,
\citet{smith_multi-radar_2016}) product along the U.S. East Coast. For the
CPR-based estimates, all available overpasses from June 2015 through the end of
the CloudSat mission are included. Owing to the wider swath of the DPR,
overpasses from the year 2022 alone provide sufficient sampling for a robust
comparison.

Because this study focuses on oceanic precipitation, only satellite pixels
located over the ocean are retained. Since the chance for beam overshoot
increases with distance from the shore, only MRMS pixels with a radar quality
index (RQI) exceeding 0.9 are retained, limiting the distance to the closest
radar to about 100 km. As RQI fields are unavailable for some earlier MRMS
periods, the mean RQI field from 2022 is used to screen the matched
observations. Following \citep{pfreundschuh2026benchmark}, the MRMS
precipitation fields are aggregated to a $0.36^\circ \times 0.36^\circ$ grid and
collocated with the CloudSat and DPR estimates using nearest-neighbor
interpolation. Panel (a) of Fig. 5 presents the mean radar quality index (RQI)
used to filter valid match-ups. Even with a minimum RQI threshold of 0.9, the
NEXRAD network provides coverage across most of the U.S. East Coast. Panels (b)
and (c) display the total number of valid precipitation retrievals extracted
within the validation domain before the RQI mask is applied. Valid CloudSat
retrievals are largely limited to ascending overpasses due to its restriction to
daytime-only operation during the validation period. A smaller number of
descending overpasses are available over the northern portion of the domain
during summer. In contrast, the substantially wider swath of the DPR results in
far more frequent observations that are distributed more uniformly across the
domain.

\section{Results}
\label{sec:results}

Below we present the three principal results from this work: Firstly, we assess
the accuracy of the GPROF-NN XPR retrieval by comparing it to GPM CMB and
CloudSat estimates from independent testing periods and a case study from the
validation period. Following this, we validate the fused GPROF-NN XPR retrieval
as well as CPR- and DPR-based precipitation estimates against independent
in-situ precipitation estimates. Finally, we compare global retrieval statistics
to existing datasets to assess the impact of the expanded reference data on the
GPROF-NN XPR retrievals.

\subsection{Retrieval Accuracy}

\begin{figure}[h]
 \centerline{\includegraphics[width=37pc]{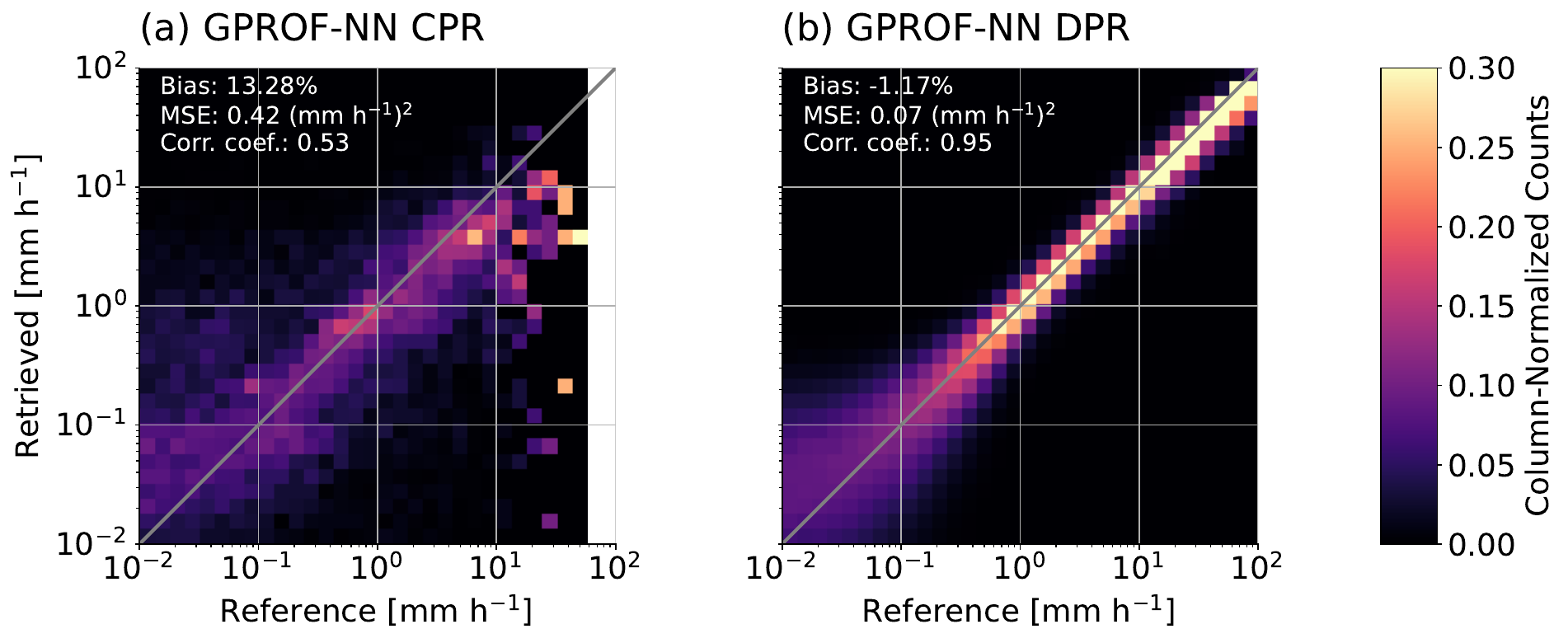}}
 \caption{
   Scatter plots of reference and retrieved precipitation for the (a) CPR-based
   and (b) the DPR-based precipitation estimates.
 }\label{fig:test_results}
\end{figure}

Figure~\ref{fig:test_results} shows scatter plots of reference and retrieved
precipitation rates for the CPR- and DPR-based outputs of the GPROF-NN XPR
retrieval. The retrieval reproduces the DPR-based estimates well, achieving low
bias and mean-squared error and high linear correlation. The accuracy of the
CPR-based estimates is substantially lower: MSE and correlation coefficient
indicate lower accuracy that is also reflected in the increased spread of the
retrieved values. In part, this can be explained by the higher resolution of the
CPR-based precipitation estimates. The training and testing datasets both use CPR
estimates at the native resolution (1.6 km), whereas the DPR-based estimates are
downsampled to the GMI resolution, which is how they are traditionally treated
in the GPROF framework. The different approach for the CPR-based estimates was
chosen due to the difficulties of assigning a unique resolution to the GMI
observations and of matching the one-dimensional CPR data to the
two-dimensional GMI footprints.

The CPR-based precipitation estimates show fair agreement with the reference
values from 0.1 to 3--4 mm h$^{-1}$ but struggle to reproduce precipitation estimates
exceeding that. However, since CloudSat estimates have to be considered very
uncertain in this range, it is possible that this is – at least in part – due to
errors in the CPR-based reference data. Additionally, there is a notable bias in
the CPR-based precipitation estimates of about 13\% which is caused by the
underestimation of the higher rain rates. Given that precipitation rates
exceeding 3--4 mm/h account for more than 10 \% of the total precipitation in the
test data, the bias may be caused by statistical fluctuations and the smaller
number of testing samples compared to the DPR-based estimates. Furthermore,
since CloudSat left the A-train in 2018, the precipitation estimates from the
test data were derived in a different orbital configuration, which can further
affect bulk precipitation statistics and thus produce biases in the test
results.

\subsection{Case Study}

\begin{figure}[h]
 \centerline{\includegraphics[width=37pc]{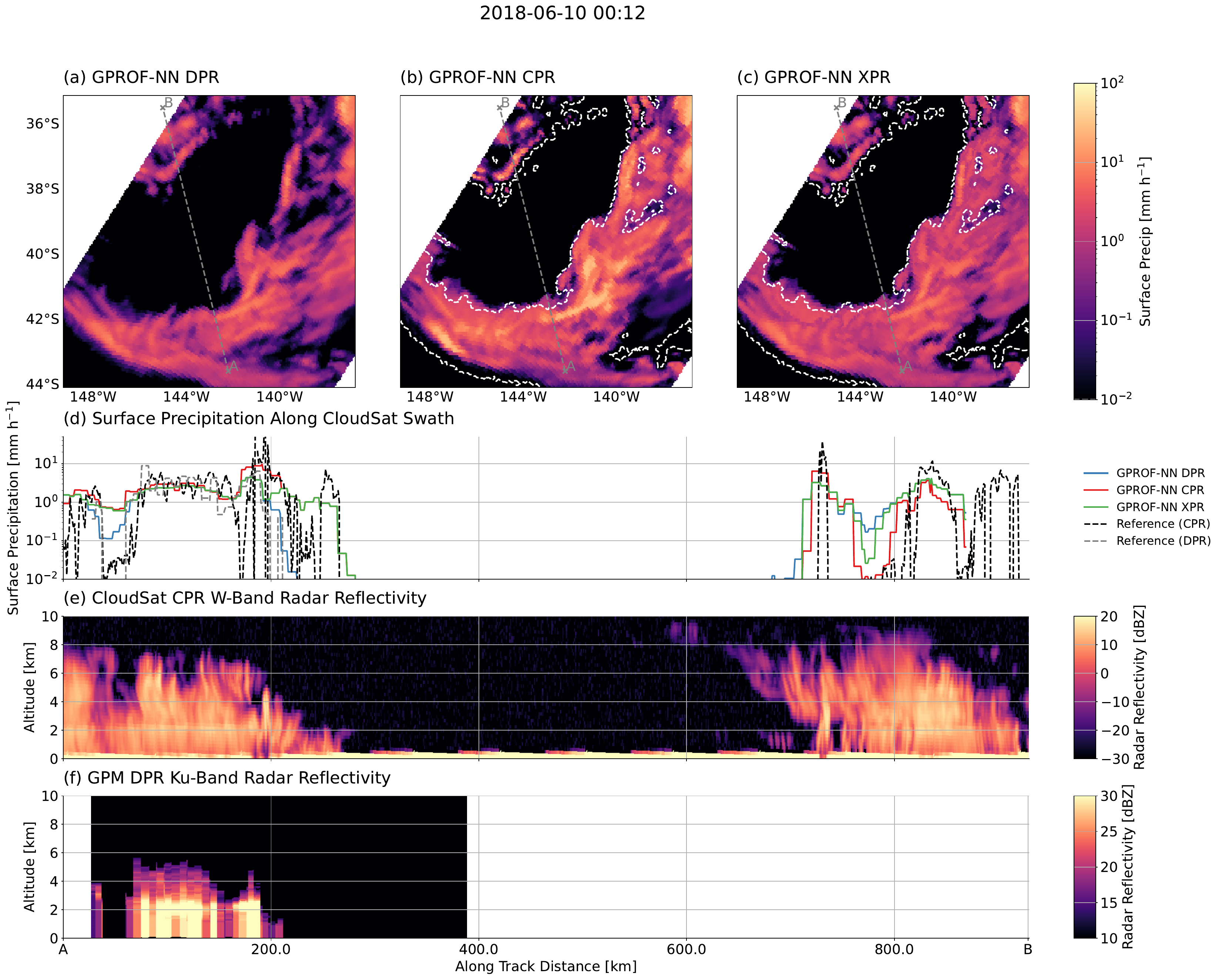}}
 \caption{
   Coincident observations of a mid-latitude cyclone by the GPM and CloudSat
   satellites. The first row of panels displays GMI-based PMW retrievals trained on
   GPM DPR precipitation (a), on CloudSat CPR precipitation (b), and the fused
   precipitation field combining the DPR- and CPR-based estimates (c ). Panels (d),
   (e), and (f) show the retrieved precipitation and the corresponding CloudSat and
   GPM radar reflectivities along the CloudSat swath.
 }\label{fig:case_study}
\end{figure}

Figure~\ref{fig:case_study} illustrates an example of the DPR- and CPR-based
precipitation estimates together with the resulting fused estimate. The case
study displays a GMI overpass of a Southern Hemisphere mid-latitude cyclone in
June 2018. A comparison of the CPR-based, DPR-based, and fused XPR precipitation
fields highlights substantial structural differences among the three products.
Although the CPR- and DPR-based estimates are generated using the same model,
they exhibit marked discrepancies. The CPR-based retrieval produces higher
precipitation intensities and a broader spatial extent of the cyclone’s
precipitation field. The fused XPR product moderates the elevated precipitation
rates evident in the CPR-based retrieval while largely preserving its enhanced
spatial coverage.

Comparison with the reference data along the CloudSat swath shows that the
GPROF-NN XPR retrieval successfully captures the shallow precipitation occurring
between 200 and 250 km along the CloudSat track that is missed by both the DPR
reference estimates and the GPROF-NN DPR retrievals. Generally, the GPROF-NN XPR
results capture the main precipitation features but do not reproduce the
small-scale variability present in the radar-based reference estimates. This
discrepancy is expected because PMW observations have coarser spatial resolution
and sampling than the radar measurements. At around 50 km and 750 km along the
swath, the GPROF-NN XPR retrieval indicates precipitation while the CPR-based
estimates suggest little or no precipitation. However, the CPR radar still
detects reflectivities extending down to the surface in these regions,
suggesting that precipitation likely reaches the surface but may be
overestimated by the PMW retrieval.

\subsection{Validation}

The results presented above demonstrate that the GPROF-NN XPR retrieval is able
to reproduce both CPR- and DPR-based precipitation estimates accurately
considering the differences in resolution and information content of the PMW
observations as well as the sensitivity ranges of the two reference datasets.
However, since neither reference dataset accurately captures precipitation
accurately across the full spectrum of precipitation intensities, it is
essential to evaluate the GPROF-NN XPR retrieval against independent
precipitation measurements to ensure it actually improves the representation of
precipitation cipitation. Below we validate both the GPROF-NN XPR retrieval as
well as the reference datasets themselves against independent precipitation
measurements.

\subsubsection{OceanRAIN}

The GPROF-NN XPR precipitation estimates are evaluated using 3410 OceanRAIN
overpasses that passed quality control and yielded valid precipitation estimates
from GPROF V07, ERA5, and GPROF-NN XPR. To provide a more granular assessment,
we stratify the analysis into three subsets: (1) all overpasses, (2)
high-latitude cases poleward of 45°, and (3) scenes with a high likelihood of
mixed or frozen precipitation. The latter classification uses the
parameterization of \citet{sims2015parameterization} with ERA5-derived wet-bulb
temperatures. Overpasses with a conditional probability of solid precipitation
exceeding 50\% are categorized as likely mixed or frozen precipitation events.

\paragraph{Precipitation Detection}

\begin{figure}[h]
 \centerline{\includegraphics[width=37pc]{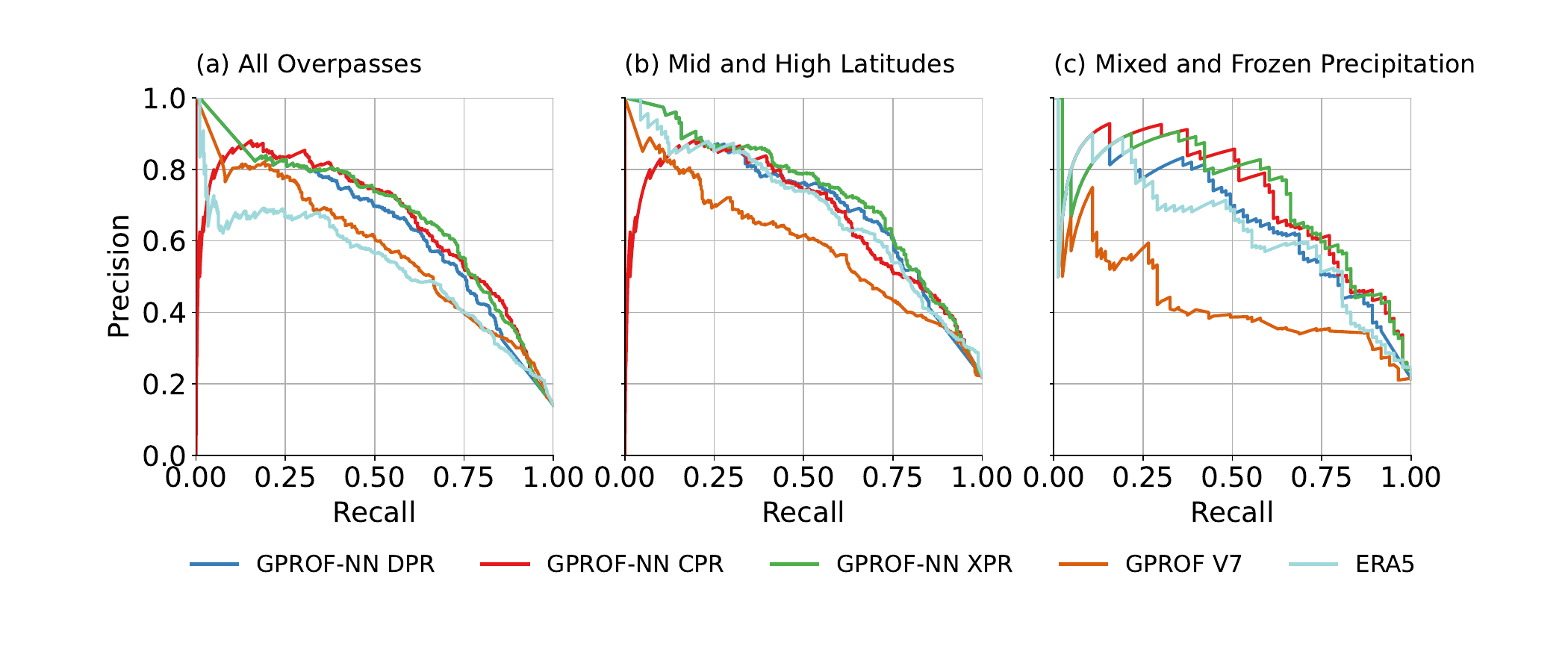}}
 \caption{
    Precipitation-detection skill evaluated against ship-borne distrometer
    measurements. Each panel shows precision–recall curves for the GPROF-NN XPR
    retrieval outputs based on CPR estimates (GPROF-NN CPR), the DPR-based
    precipitation estimates (GPROF-NN DPR), the fused estimates (GPROF-NN XPR) and
    the GPROF V07 and ERA5 baselines. c Panels (a) – (c ) show precipitation
    estimates for all overpasses, overpasses for latitudes poleward of 45°, and for
    mixed and frozen precipitation, respectively.
 }\label{fig:ocean_rain_detection}
\end{figure}

We first assess the detection capability of the three GPROF-NN retrievals as
well as GPROF V07 and ERA5 baselines using precision-recall curves. The
precision recall curves display the trade off between precision, i.e., the
probability of a detection being correct, and the recall, i.e., the fraction of
total precipitation events detected, as the probability threshold used to decide
whether a pixel is raining or not is varied. The more skillful the detection, the
closer the curve gets to a precision and recall value of one. All GPROF
retrievals provide a probability of precipitation representing the estimated
probability of the corresponding pixel containing precipitation. Since ERA5
doesn’t provide a dedicated probability field for precipitation occurrence, we
use the precipitation rate as detection criterion. An overpass is defined to be
precipitating when the half-hour average distrometer precipitation exceeds 0.01
mm/h.

Figure~\ref{fig:ocean_rain_detection} displays the precipitation detection
accuracy for all overpasses, overpasses at high latitudes, and mixed and frozen
precipitation. Considering all overpasses, the CloudSat-based light
precipitation and the fused GPROF-NN XPR results yield the highest detection
accuracy. This is consistent with the CloudSat-based precipitation estimates
providing higher sensitivity to light precipitation that is missed by the GPM
DPR reference data. The retrieval based solely on GPM DPR data has slightly
lower detection skill but remains clearly more sensitive than the GPROF V07 and
ERA5 baselines.

Somewhat surprisingly, the detection accuracy of the DPR-based estimates exceeds
that of the CPR-based estimates at mid and high latitudes. This seems to
indicate that the CPR-based GPROF-NN estimates miss certain precipitation events
that are detected by the DPR-based estimates but it is not immediately clear for
which precipitation processes this would occur. The GPROF-NN XPR achieves the
highest precipitation detection skill exceeding that of both the CPR- and the
DPR-based retrievals.

Although the results for mixed and frozen precipitation are overall noisier, the
CPR-based and fused precipitation estimates clearly achieve the best detection
accuracy. The DPR-based retrieval performs worse and is closely followed by
ERA5. The GPROF V07 retrieval performs substantially worse with the best
precision and recall values being about half of the GPROF-NN XPR retrieval.

Table~\ref{tab:detection_metrics} evaluates the detection skill of all
retrievals that provide a probability of precipitation using a probability
threshold of 0.5. This evaluation confirms that the GPROF-NN XPR retrieval
yields the best detection skill across all precipitation regimes. In particular,
the GPROF-NN XPR retrieval improves the detection skill compared to GPROF-NN DPR
by 48 \%, 26 \%, and 42 \% for all overpasses, mid and high latitudes
overpasses, and frozen precipitation, respectively.

\begin{table}[t]
\centering
\caption{Precipitation detection metrics (POD, FAR, CSI) for different retrieval methods evaluated against shipborne distrometer measurements across precipitation regimes.}
\label{tab:detection_metrics}
\begin{tabular}{lccccccccc}
\toprule
& \multicolumn{3}{c}{All Overpasses} 
& \multicolumn{3}{c}{Mid and High Latitudes} 
& \multicolumn{3}{c}{Frozen Precipitation} \\
\cmidrule(lr){2-4} \cmidrule(lr){5-7} \cmidrule(lr){8-10}
Method 
& POD & FAR & CSI 
& POD & FAR & CSI 
& POD & FAR & CSI \\
\midrule
GPROF V07    & 0.37 & 0.31 & 0.32 & 0.29 & 0.28 & 0.26 & 0.11 & 0.25 & 0.10 \\
GPROF-NN DPR & 0.85 & 0.65 & 0.33 & 0.85 & 0.56 & 0.41 & 0.90 & 0.63 & 0.36 \\
GPROF-NN CPR & 0.44 & 0.23 & 0.39 & 0.45 & 0.23 & 0.40 & 0.66 & 0.35 & 0.49 \\
GPROF-NN XPR & 0.67 & 0.35 & 0.49 & 0.65 & 0.28 & 0.52 & 0.71 & 0.36 & 0.51 \\
\bottomrule
\end{tabular}
\end{table}

\paragraph{Precipitation Estimation}

\begin{figure}[h]
 \centerline{\includegraphics[width=37pc]{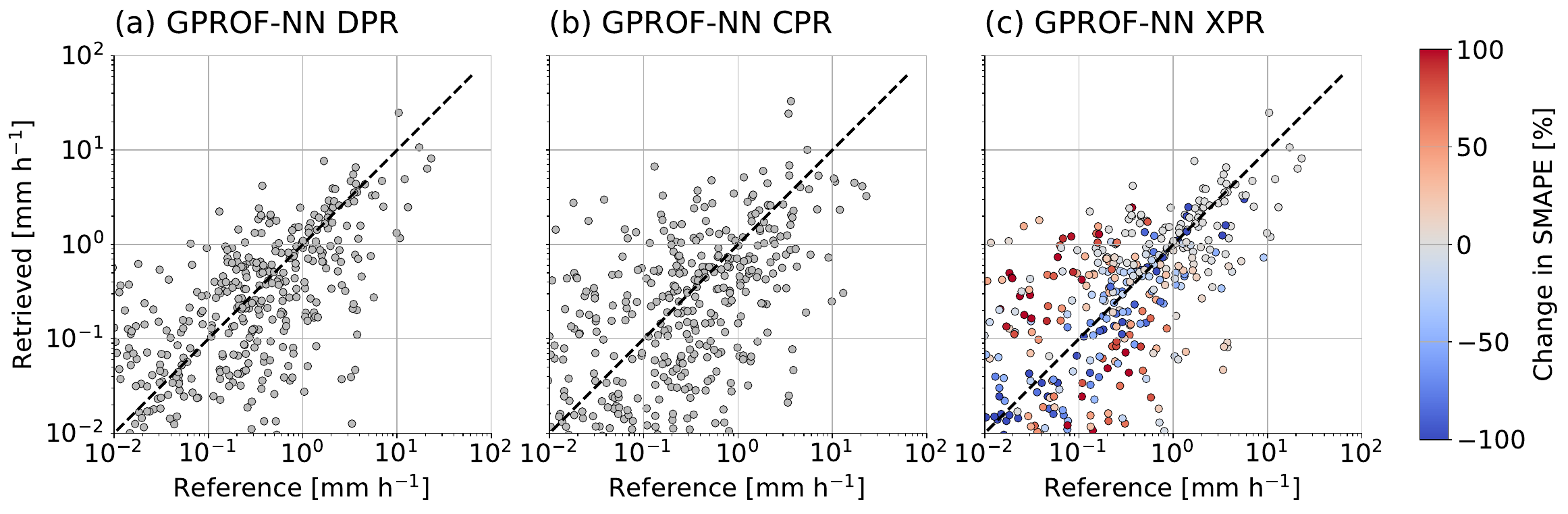}}
 \caption{
 OceanRain measurements and matches GPROF-NN precipitation retrievals. Panels
 (a) – (c) display scatter plots of OceanRAIN reference precipitation and
 corresponding precipitation retrieved from the GPROF-NN DPR, GPROF-NN CPR, and
 GPROF-NN XPR retrievals. Shading in panel (c) indicates the change in the
 Symmetric Mean Percentage Error resulting from the merging of the DPR- and CPR-
 based estimates relative to the GPROF-NN DPR baseline.
 }\label{fig:ocean_rain_scatter}
\end{figure}

Figure~\ref{fig:ocean_rain_scatter} assesses the quantitative precipitation
estimates of the GPROF-NN retrievals with respect to the OceanRAIN reference.
The GPROF-NN DPR retrieval achieves higher correlation with the OceanRAIN
measurements than the CPR-based estimates, which exhibit higher spread at both
low- and high-precipitation rates. To assess the impact of the
precipitation-rate fusion, panel (c) displays the relative change in the
symmetric mean percentage error (SMAPE) resulting from the combining of the DPR-
and CPR-based estimates compared to the DPR baseline. While the merging
decreases the SMAPE for a significant number of samples ending up close to the
diagonal, the are a notable amount of samples for which the SMAPE is increased.
Most of these correspond to the merging leading to an overestimation of the
precipitation. Since the coloring indicates the increase in the relative error,
the largest increases in error occur at small reference precipitation rates as
the relative error is more sensitive in these regimes. For a smaller number of
red samples below the diagonal, the merging with the CPR estimates erroneously
reduces the final precipitation rate.

\begin{figure}[h]
 \centerline{\includegraphics[width=37pc]{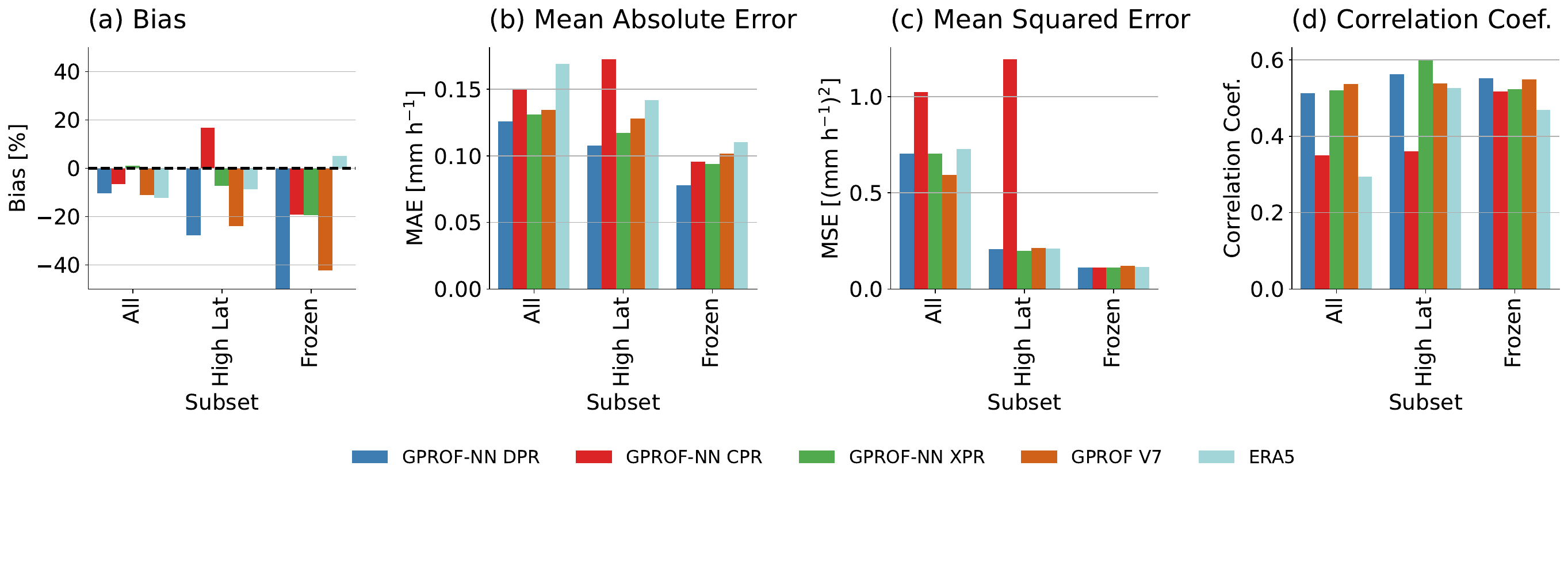}}
 \caption{
Statistical evaluation of the GPROF-NN and baseline datasets against shipborne
in-situ measurements from the OceanRAIN dataset. Panels (a) – (d) display bias,
mean absolute error, mean squared error, and linear correlation coefficient for
GMI overpasses over the OceanRAIN vessels. Results are reported independently
for all overpasses (All), overpasses poleward of 45 deg (High Lat) and
overpasses with an estimated fraction of frozen precipitation exceeding 50
\% (Frozen).
 }\label{fig:ocean_rain_stats}
\end{figure}

A statistical evaluation of the OceanRAIN overpasses is presented in Fig.~\ref{fig:ocean_rain_stats}.
Evaluated over all overpasses, both the DPR- and CPR-based estimates are biased
low. The bias of the DPR-based estimates is around 10 \%, while the DPR-based
estimates are biased low by 7 \%. Fusing the two estimates leads to a reduction
in bias to a positive bias of about 1 \%. At mid- and high latitudes, the low
bias of the DPR estimates is enhanced while the CPR estimates are biased high
with 16 \%. The fused precipitation estimates are biased low by about 7 \%. For
frozen precipitation, the bias in the DPR-based estimates increases to 51 \%
while the CPR-based estimates are low by 20 \%. The bias in the fused GPROF-NN
XPR precipitation estimates is -20 \%, similar to that of the CPR-based
estimates. The fusion of the DPR- and CPR- based estimates thus helps to reduce
systematic underestimation of the DPR-based retrieval. Furthermore, the biases
are smaller than for the GPROF V07 baseline as well as the ERA5 baseline except
for frozen precipitation.

In terms of instantaneous retrieval error, the CPR-based estimates have higher
errors than the DPR-based estimates, particularly at mid to high latitudes. The
fused estimates dampen these errors to some extent but remain higher or at the
level of the DPR-based estimates. The GPROF-NN XPR retrieval generally achieves
lower mean absolute and mean squared error than the ERA5 baseline as well as
higher correlation. The GPROF V07 baseline has smaller mean squared error and
higher correlation coefficient for all overpasses and higher correlation for
frozen precipitation. However, comparing the GPROF estimates to the actual
DPR-based reference estimates using the 1028 OceanRAIN overpasses with valid
2BCMB reference precipitation estimates, the GPROF-NN DPR estimates achieve a
MSE of 0.26 (mm h$^{-1}$)$^2$ and a linear correlation coefficient of 0.84
compared to 0.32 (mm h$^{-1}$)$^2$ and a linear correlation of 0.8 for the GPROF
V07 baseline. This indicates that these small accuracy differences are likely
due to either statistical noise or uncertainties in the validation data.

\subsubsection{Validation of CPR- and DPR-based Precipitation Estimates}

The validation against OceanRAIN shipborne distrometer measurements has shown
that the fusion of CPR- and DPR-based precipitation estimates improves the
precipitation detection skill and reduces systematic underestimation but
increases the random errors in the precipitation estimates. Assessment of the
CPR- and DPR-based estimates showed that these higher errors originate from the
CPR-based estimates, which exhibit significantly larger random errors than the
DPR-based estimates. Below we therefore examine CPR- and DPR-based precipitation
errors against independent estimates from ground-based radars.

\begin{figure}[h]
 \centerline{\includegraphics[width=37pc]{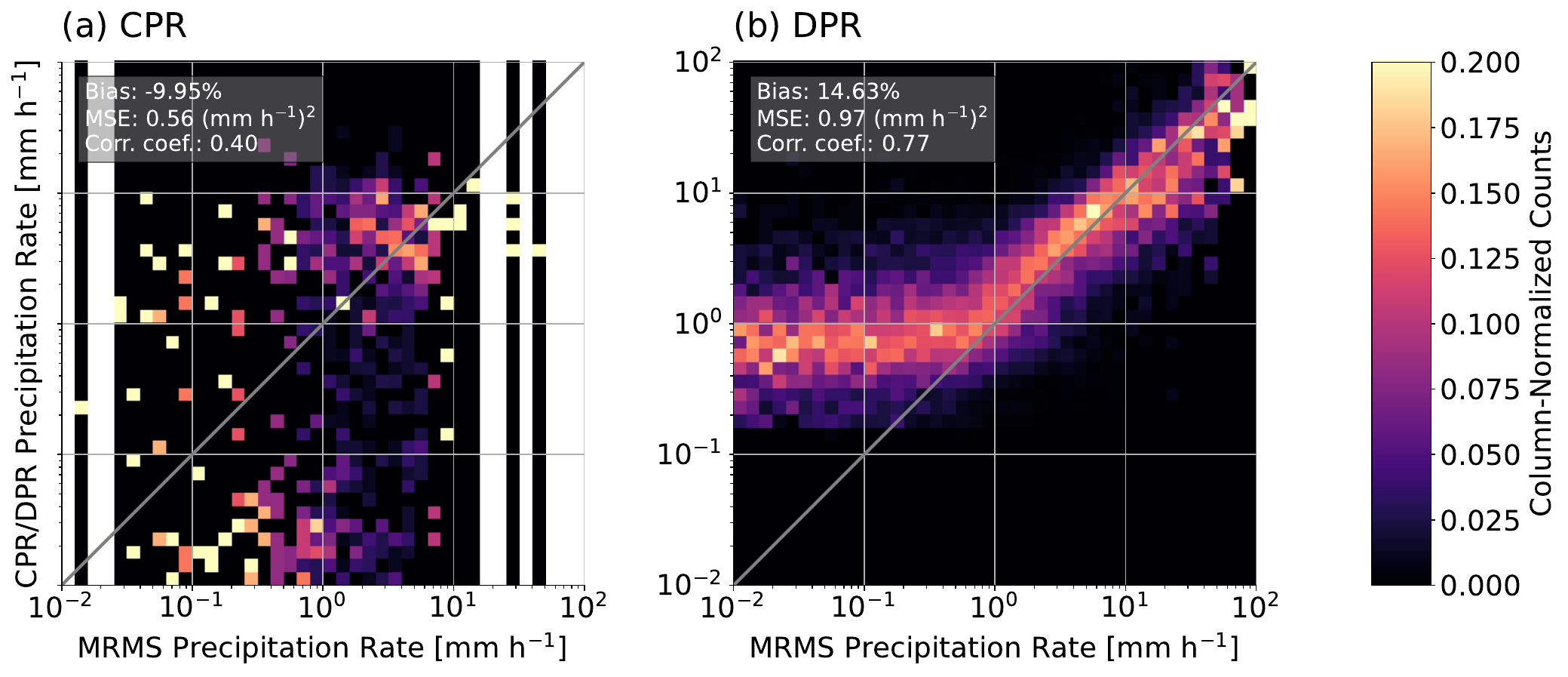}}
 \caption{
Scatter plots of precipitation estimates from the (a) CPR- and (b) DPR-based
reference precipitation estimates compared to independent estimates from
ground-based radar.
 }\label{fig:mrms_scatter}
\end{figure}

Figure~\ref{fig:mrms_scatter} displays scatter plots of the CPR- and DPR-based
precipitation estimates compared to ground-based radar measurements of the US
East Coast. The CPR-based precipitation estimates exhibit very high scatter
compared to the MRMS-based measurements particularly for MRMS precipitation
rates up to 1 mm/h. This is surprising given that the CloudSat estimates are
expected to be more reliable at low precipitation rates, where the radar
observations are less affected by attenuation. Nonetheless, the estimates
achieve a correlation of 0.4 likely due to the better agreement for MRMS
precipitation rates exceeding 1 mm h$^{-1}$. The scatter plot indicates a
distinct bi-modal behavior for MRMS precipitation rates less than 1 mm h$^{-1}$,
where the retrieval tends to either heavly overestimate or underestimate the
MRMS precipitation rates but almost never matches them.

The DPR-based estimates show better agreement with the MRMS estimates. Although
the DPR estimates strongly overestimate any MRMS estimates below 0.5 mm/h, which
is a result of the limited sensitivity of the DPR, the show very good agreement
with the MRMS-based estimates for precipitation rates from 5--10 mm/h, after
which they start to exhibit a tendency towards under-estimation.

\begin{figure}[h]
 \centerline{\includegraphics[width=37pc]{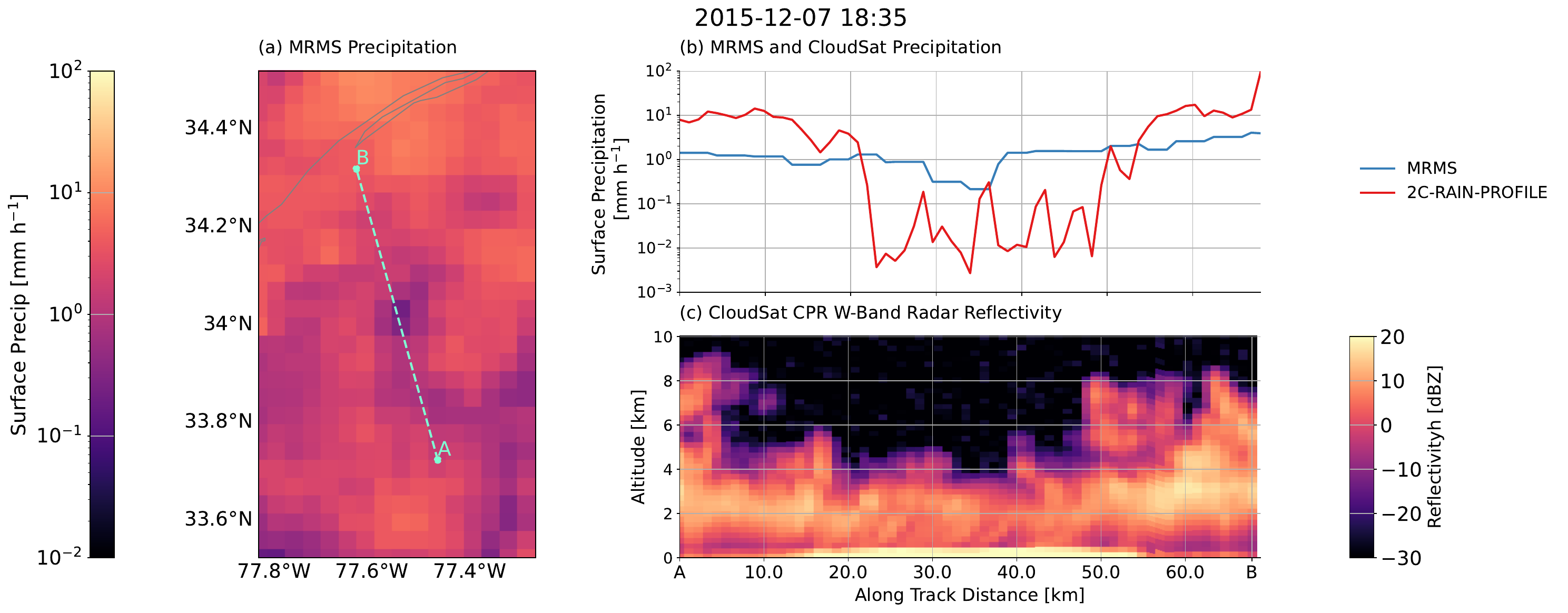}}
 \caption{
   Comparison of collocated CPR and MRMS precipitation estimates. Panel (a)
   displays the MRMS precipitation field and the ground track of the CPR
   observations. Panels (b) and (c) display the MRMS and CPR precipiation rates
   and the CPR radar reflectivity along the track, respectively.
 }\label{fig:mrms_case_study}
\end{figure}

To illustrate the likely failure mode of the CPR precipitation retrievals,
Figure~\ref{fig:mrms_case_study} shows one of the CPR overpasses from the
validation dataset. While the MRMS precipitation measurements indicate spatially
homogeneous precipitation with precipitation rates around 1 mm h$^{-1}$ throughout most of the
scene, the CPR-based precipitation oscillate between 0.01 and 10 mm h$^{-1}$.
The principal change in CPR radar reflectivities during the transition from 10
mm h$^{-1}$ to 0.01 mm h$^{-1}$ is a strong increase in the surface backscatter
while the above-surface reflectivities remain relatively similar.
 While limited, this evidence suggest that the errors in the CPR-based
 precipitation estimates may be related to the interpretation of the surface
 backscatter. Specifically in this scene where the surface backscatter
 increases, the estimate of the path integrated attenuation and column
 integrated precipitation water content decrease. One plausible source of
 uncertainty in this case is a localized change in the surface roughness not
 accounted for in the CloudSat algorithms.

\begin{figure}[h]
 \centerline{\includegraphics[width=37pc]{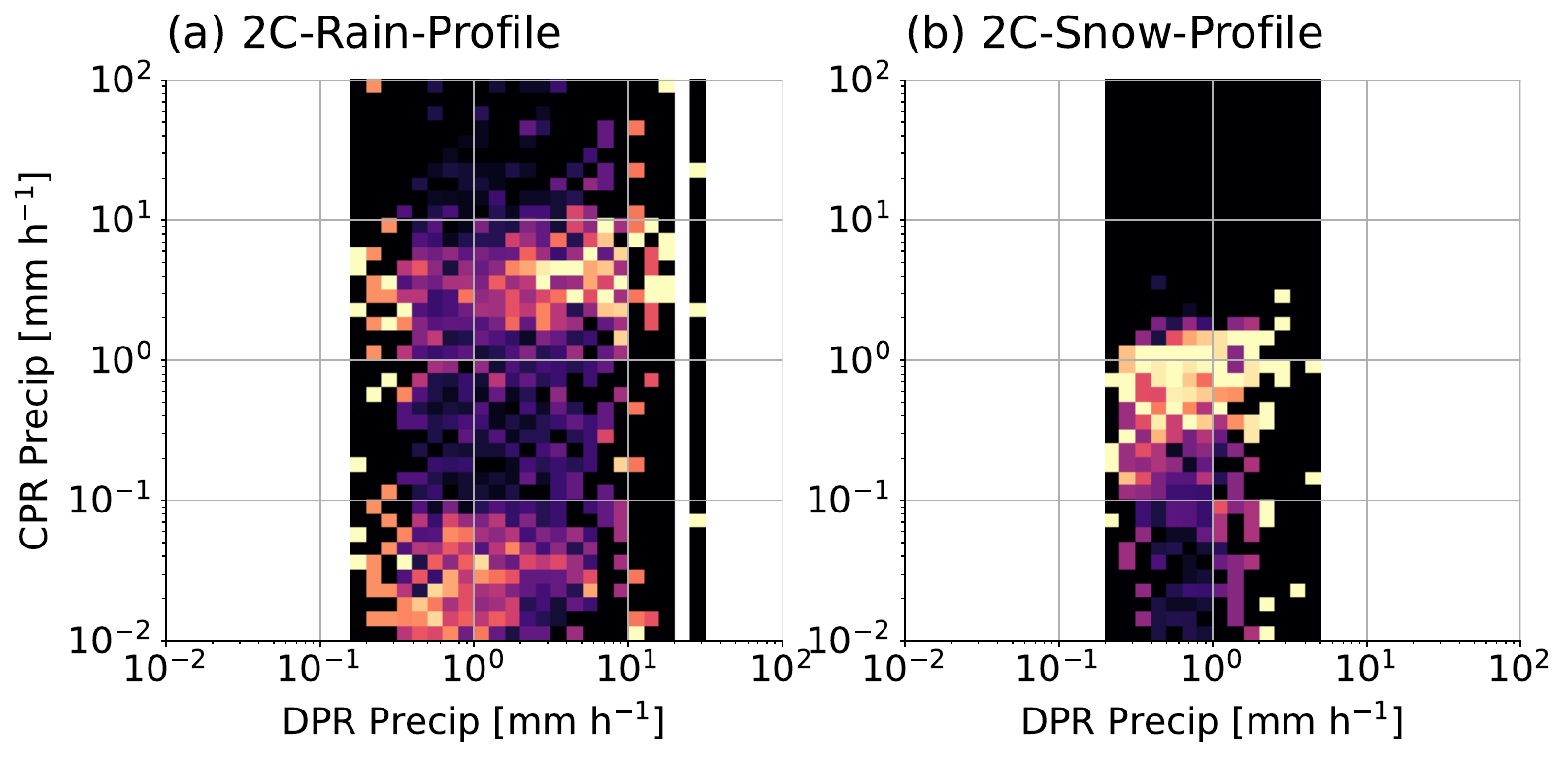}}
 \caption{
Scatter plot comparing CPR- and DPR- based precipitation estimates. Panel (a)
compares the DPR-based precipitation estimates from the GPM 2BCMB product with
estimates of the CloudSat 2C-Rain-Profile product. Panel (b) compares the GPM
estimates with snowfall estimates from the 2C-Snow-Profile product.
 }\label{fig:cpr_cmb_scatter}
\end{figure}

The comparison against the independent precipitation estimates from MRMS
indicates that the CPR-based precipitation estimates do not yield an accurate
quantitative estimate of light precipitation. In particular, the retrieval seems
to exhibit a tendency to either strongly underestimate or strongly overestimate
MRMS precipitation rates within 0.1 to 1 mm/h. As shown in
Fig.~\ref{fig:cpr_cmb_scatter}, this behavior is also observed when the CPR
estimates are directly compared to the DPR-based estimates. Although one would
expect the CPR and DPR estimates to agree across precipitation estimates from
0.5 to 1 mm h$^{-1}$, the scatter plot exhibits similar bi-modal behavior as for
the validation against MRMS. The results in Fig. 11 further show that the
inconsistency between the estimates is largely caused by the precipitation
estimates from the 2C-Rain-Profile product. The snowfall estimates from the
2C-Snow-Profile product show generally better consistency particularly for
estimates around 1 mm h$^{-1}$.

\subsection{Global Precipitation Distributions}

Although the fusing of CPR- and DPR- based precipitation estimates does not
increase the accuracy of individual precipitation estimates, it has shown clear
improvements in precipitation detection and reductions of climatological biases
compared to shipborne distrometer measurements. This indicates that the fused
GPROF-NN XPR estimates better represent light and frozen precipitation. Below we
characterize the impact of fusing CPR and DPR-based precipitation estimates on
global precipitation distributions.

\begin{figure}[h]
 \centerline{\includegraphics[width=37pc]{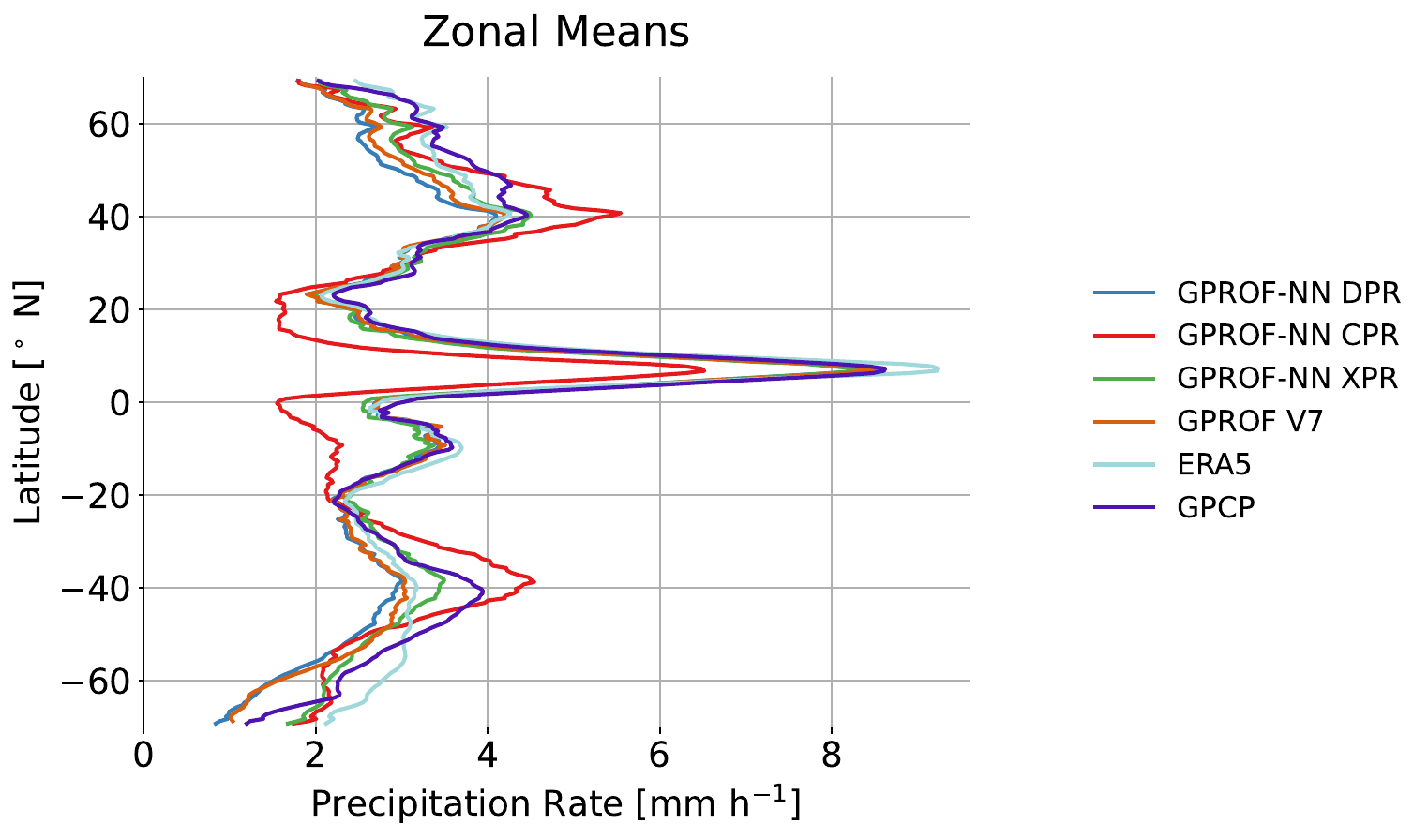}}
 \caption{
Zonal means of oceanic precipitation estimates. The figure compares estimates
from the CPR-based (GPROF-NN CPR), the DPR-based (GPROF-NN DPR), and from the
fused precipitation estimates (GPROF-NN XPR) to reference estimates from GPROF V07, ERA5,
and GPCP.
 }\label{fig:zonal_means_all}
\end{figure}

Figure~\ref{fig:zonal_means_all} displays the zonal distribution of ocean
precipitation as retrieved by the GPROF-NN CPR, DPR, and XPR retrievals. The
CPR-based retrieval yields higher precipitation rates poleward of the
mid-latitude storm tracks but severely underestimates precipitation in the
tropics and extratropics compared to the estimates derived from DPR-based
retrieval. The fused estimates track the GPM DPR-based estimates within
30$^\circ$ S -- 30$^\circ$ N. Poleward of that, the GPROF-NN XPR estimates
mostly fall between the DPR and CPR based estimates approximating the CPR-based
estimates at high latitudes. The increased precipitation at mid and high
latitudes moves the zonal profiles of the GPROF-NN XPR retrieval closer to those
of of GPCP but remains lower than them at all latitudes. The GPROF-NN XPR
retrieval is slightly higher than ERA5 in the storm tracks but falls below it at
high latitudes.

\begin{figure}[hp]
 \centerline{\includegraphics[width=37pc]{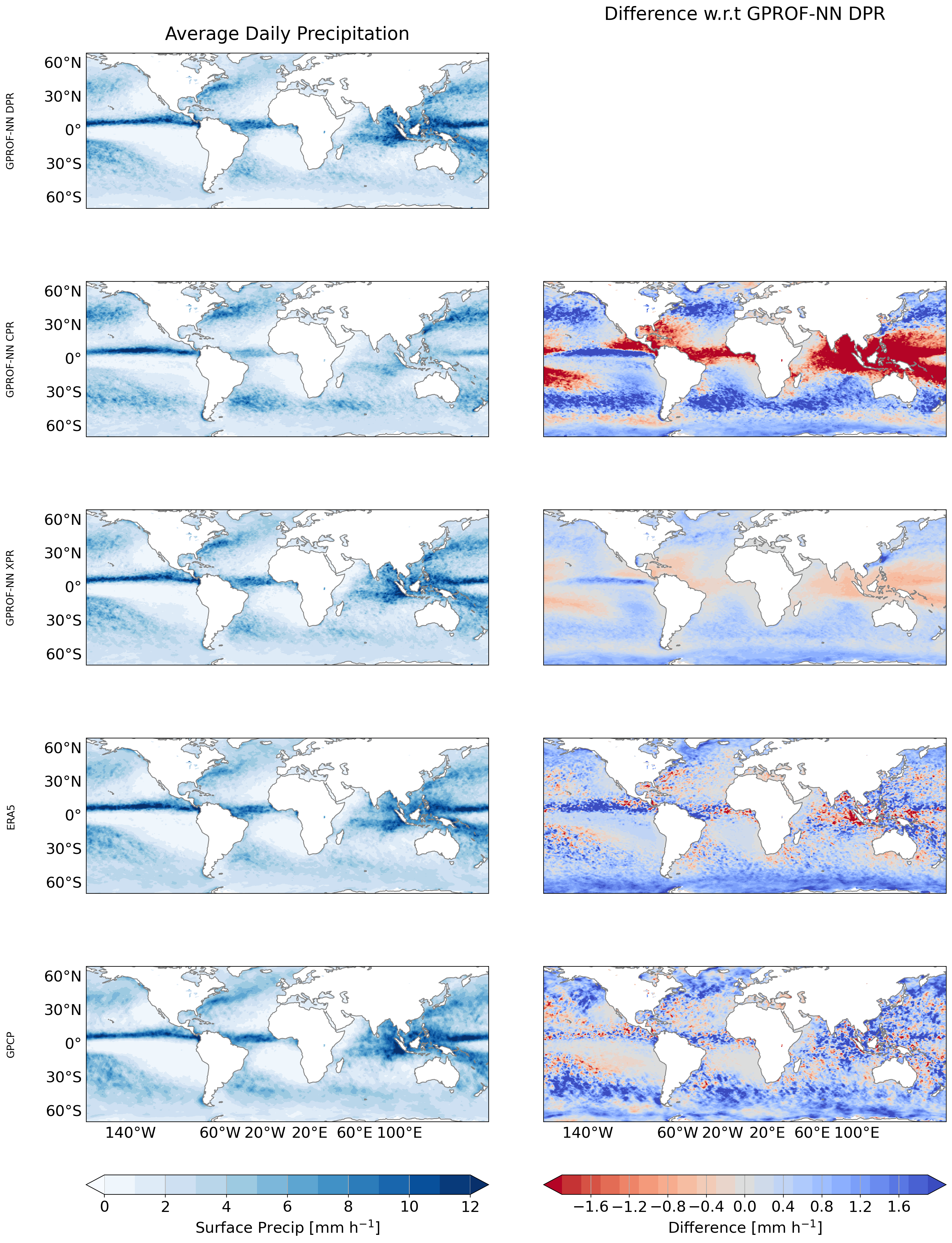}}
 \caption{
Global distributions of mean daily precipitation accumulations from the DPR-based estimates (GPROF-NN DPR), the CPR-based estimates (GPROF-NN CPR), and the merged estimates (GPROF-NN XPR) and the ERA5 and GPCP datasets.
 }\label{fig:global_means_all}
\end{figure}

The corresponding global precipitation distributions are displayed in
Fig.~\ref{fig:global_means_all}. Compared to the DPR-based retrievals, the
CPR-based estimates add more than 2 mm per day over the storm tracks and
high-latitude oceans. The fused XPR estimates add precipitation over the storm
tracks and high latitudes albeit less than the GPROF-NN CPR retrieval alone.
The fused precipitation estimates slightly decrease precipitation over the
Indian ocean and West Pacific but increase it over the eastern Pacific.

ERA5 exhibits higher precipitation rates than the DPR-based reference retrievals
almost everywhere. However, pockets of decreasing precipitation are visible in
the Indian Ocean and the West Pacific roughly matching areas of decreasing
precipitation in the GPROF-NN XPR results. GPCP precipitation rates are also
globally higher with scattered regions of decreasing precipitation.

Although the increased precipitation in the GPROF-NN XPR retrieval brings the
zonal means closer to those of ERA5 and GPCP, the analysis of the global
distribution highlights important differences between where the precipiation is
added. In contrast to ERA5 and GPCP, the GPROF-NN XPR retrieval adds
precipitation over the subtropical oceans off the west coasts of North and South
America, Africa, and Australia. In addition to this, the GPROF-NN XPR slightly
decreases precipitation over the tropical oceans east of the Americas. This may
indicate that the fused retrieval is in fact sensitive to precipitation regimes
that may be missed by the other datasets. However, given the inaccuracies in the
CPR-based reference estimates, further validation is  required to confirm
this.

\section{Summary and Conclusions}
\label{sec:summary}

This work presents GPROF-NN XPR, a novel precipitation retrieval designed to
overcome limitations of current global PMW precipitation estimates imposed by
the shortcomings of the precipitation estimates used as targets for empirical
retrieval algorithms. GPROF-NN XPR retrieval combines dedicated estimates of
light precipitation based on reference data from the CloudSat CPR with
moderate-to-light precipitation estimates based on reference data from the GPM
DPR. A simple fusion scheme is proposed to combine the two precipitation
estimates into a single fused precipitation estimate. This novel approach
extends previous efforts aiming to use CloudSat precipitation rates to retrieve
light and frozen precipitation from PMW sensors \citep{rahimi2025advancing,
  eastman2019}.

Validation of the fused GPROF-NN XPR precipitation estimates shows improved
precipitation detection and a reduction of the systematic underestimation that
affects conventional retrievals, particularly at high latitudes and for frozen
precipitation. These results indicate that GMI PMW observations can help bridge
the current sensitivity gap between radar-based precipitation measurements from
the CloudSat and GPM missions. By providing estimates that better represent
preciptiation across regimes, the GPROF-NN XPR retrieval has the potential to
improve global precipitation datasets. In particular, it may reduce the need for
climatological corrections that suppress monthly and interannual variability.

The improved represntation of precipitation across precipitation regimes in the
PMW retrievals, also demonstrates that there is potential to improve synergistic
active–passive precipitation retrievals using GMI observations. The DPR-based
estimates used here are derived from the GPM 2BCMB product, which already
combines radar and GMI observations. However, the current algorithm only
retrieves precipitation where the radar observations indicate hydrometeor
backscatter, which limits the retrievals to the low sensitivity of the DPR.
The results presented here demonstrate that GMI observations alone can identify
precipitation that is not detected by the DPR, suggesting that current combined
radar--radiometer retrievals do not yet fully exploit the synergistic potential
of the active and passive microwave observations provided by the GPM satellite.

Although the fused precipitation estimates improve detection and reduce
climatological biases, the fusion increases random retrieval errors compared to
the DPR-based estimates. Validation of the CPR-based precipitation estimates
from the 2C-Rain-Profile product provide more sensitive but generally less
accurate quantitative precipitation estimates than the DPR-based reference. The
comparison against both ground-based radar and against DPR show large spread at
light precipitation rates with a bi-modal tendency to either strongly
underestimate or overestimate precipitation by an order of magnitude. The
bi-modal behavior of the 2C-Rain-Profile was already noted in the first
publication describing the retrieval \citep{lebsock_2011} but apparently has not
been further investigated since. Since 2C-Rain-Profile estimates are used to
calibrate principal global precipitation products including IMERG and GPCP
\citep{behrangi_new_2020}, there is a need to revisit these precipitation
estimates and ensure consistency with other radar-based precipitation estimates
across the shared range of sensitivity.

Despite teh uncertainties in the quantitative precipitation estimates from the
CPR, our extensive validation of the GPROF-NN XPR estimates against direct,
shipborne measuremens demonstrates the benefits of using CPR-based precipitation
estimates to improve precipitation detection and the climatological
representation of high-latitude and frozen precipitation. The GPROF-NN XPR
retrieval will therefore be used as an auxiliary retrieval to add light
precipitation to the DPR-based reference data that will be used to train the
retrieval algorithms for the next operational version of the GPM PMW retrieval
algorithms.

While the GPROF-NN XPR retrieval constitutes an additional step towards improved
global estimates of oceanic precipitation from the GPM mission, this work also
highlights remaining challenges of current global precipitation products. Since
CPR- and DPR-based are currently used in combination to calibrate global
precipitation products, it should be ensured that these products are consistent
across the shared sensitivity range of CPR and DPR. Moreover, our results
indicate that current synergistic active–passive precipitation retrievals
underutilize passive observations and could likely be improved to overcome
limitations of current active sensors.

\clearpage
%%%%%%%%%%%%%%%%%%%%%%%%%%%%%%%%%%%%%%%%%%%%%%%%%%%%%%%%%%%%%%%%%%%%%
% ACKNOWLEDGMENTS
%%%%%%%%%%%%%%%%%%%%%%%%%%%%%%%%%%%%%%%%%%%%%%%%%%%%%%%%%%%%%%%%%%%%%
\acknowledgments

The work of SP and CDK was supported by NASA grant 80NSSC22K0604.

The authors thank Mathew Lebsock for valuable feedback on the validation of the
CloudSat precipitation estimates.

%%%%%%%%%%%%%%%%%%%%%%%%%%%%%%%%%%%%%%%%%%%%%%%%%%%%%%%%%%%%%%%%%%%%%
% DATA AVAILABILITY STATEMENT
%%%%%%%%%%%%%%%%%%%%%%%%%%%%%%%%%%%%%%%%%%%%%%%%%%%%%%%%%%%%%%%%%%%%%
% 
%
\datastatement

The 2BCMB can be downloaded from \citet{cmb_data}.

The CloudSat 2C-Precip-Column, 2C-Rain-Column, and 2C-Snow-Column data are available from \citet{cloudsat_data}.

The OceanRAIN dataset is available from \citet{ocean_rain_data}.

The MRMS data can be downloaded from \citet{mrms_data}.

%%%%%%%%%%%%%%%%%%%%%%%%%%%%%%%%%%%%%%%%%%%%%%%%%%%%%%%%%%%%%%%%%%%%%
% APPENDIXES
%%%%%%%%%%%%%%%%%%%%%%%%%%%%%%%%%%%%%%%%%%%%%%%%%%%%%%%%%%%%%%%%%%%%%

%% If only one appendix, use

%\appendix

%% If more than one appendix, use \appendix[<letter>], e.g.,

%\appendix[A] 

%% Appendix title is necessary! For appendix title:

%\appendixtitle{Title of Appendix}

%%% Appendix section numbering (note, skip \section and begin with \subsection)
%
% \subsection{First primary heading}

% \subsubsection{First secondary heading}

% \paragraph{First tertiary heading}

%%%%%%%%%%%%%%%%%%%%%%%%%%%%%%%%%%%%%%%%%%%%%%%%%%%%%%%%%%%%%%%%%%%%%
% REFERENCES
%%%%%%%%%%%%%%%%%%%%%%%%%%%%%%%%%%%%%%%%%%%%%%%%%%%%%%%%%%%%%%%%%%%%%

\bibliographystyle{ametsocV6}
\bibliography{references}

\end{document}